\documentclass[conference]{IEEEtran}
\IEEEoverridecommandlockouts
\usepackage{cite}
\usepackage{amsmath,amssymb,amsfonts}
\usepackage{algorithmic}
\usepackage{graphicx}
\usepackage{textcomp}
\usepackage{xcolor}
\usepackage{textcomp}
\def\BibTeX{{\rm B\kern-.05em{\sc i\kern-.025em b}\kern-.08em
    T\kern-.1667em\lower.7ex\hbox{E}\kern-.125emX}}
    
\def\Xint#1{\mathchoice
   {\XXint\displaystyle\textstyle{#1}}%
   {\XXint\textstyle\scriptstyle{#1}}%
   {\XXint\scriptstyle\scriptscriptstyle{#1}}%
   {\XXint\scriptscriptstyle\scriptscriptstyle{#1}}%
   \!\int}
\def\XXint#1#2#3{{\setbox0=\hbox{$#1{#2#3}{\int}$}
     \vcenter{\hbox{$#2#3$}}\kern-.5\wd0}}

\def\dashint{\Xint-}
\def\dashiint{\dashint\hspace{-0.2cm}\dashint}
\newcommand{\rpos}{\mathbf{r}}
\newcommand{\rposp}{\mathbf{r}'}

\begin{document}

\title{Geometrical MoM formulation for eigenmode analysis
\thanks{\copyright 2019 IEEE. Personal use of this material is permitted. Permission from IEEE must be obtained for all other uses, in any current or future media, including reprinting/republishing this material for advertising or promotional purposes, creating new collective works, for resale or redistribution to servers or lists, or reuse of any copyrighted component of this work in other works.}}

\author{\IEEEauthorblockN{Denis Tihon}
\IEEEauthorblockA{\textit{ICTEAM institute} \\
\textit{Universit\'{e} catholique de Louvain}\\
Louvain-la-Neuve, Belgium \\
denis.tihon@uclouvain.be}
\and
\IEEEauthorblockN{Christophe Craeye}
\IEEEauthorblockA{\textit{ICTEAM institute} \\
\textit{Universit\'{e} catholique de Louvain}\\
Louvain-la-Neuve, Belgium \\
christophe.craeye@uclouvain.be}
}

\maketitle

\begin{abstract}
The resonant frequencies of a structure and the associated field distributions are generally determined by solving a non-linear eigenvalue problem. Using frequency-domain solvers, the response of the structure needs to be evaluated at many different frequencies in order to solve the non-linear problem. Moreover, these frequencies may be complex, the inverse of the imaginary part physically corresponding to the e-folding time of the energy of the mode. 
In this paper, we propose to use the so-called ``Geometrical Method of Moments" (GMoM) to accelerate the computation of the resonant frequencies of a structure using the Method of Moments. First, purely geometrical reaction integrals are precomputed, which do not depend on the frequency nor material parameters. Then, by summing these terms with proper weights, the impedance matrix can be obtained for any complex frequency. This method easily accommodates for dispersive materials provided that the permittivity and permeability of the material can be extrapolated to complex frequencies.
\end{abstract}

\begin{IEEEkeywords}
Method of Moments, Geometrical MoM, eigenmode analysis
\end{IEEEkeywords}

\section{Introduction}
In some cases, the peculiar behaviour of a structure can be accurately modelled using few resonant modes. For example, in \cite{1}, the authors explain a Fano dip in the absorption spectrum of a Dolmen structure as the combination of two resonant modes happening at the same frequency. Similarly, in that same paper, the evolution of the power absorption of a plasmonic nanowire with frequency is explained using few complex resonant frequencies, the width of the absorption peaks being directly related to the imaginary part of the complex frequency. 

Such complex resonant frequencies can be also be used for design purposes. Indeed, in some applications, it is crucial that the device that is being designed resonate at a precise frequency. It is for example the case in MRI, where the frequency of operation is fixed by the proton precession and the antennas should be designed accordingly \cite{7}.

The determination of the resonant frequency of a structure can be carried out using spectral methods \cite{1,3,9, 10, 12}. Using such methods, the problem is non-linear requiring to simulate the structure for several complex frequencies. Using integral equation methods \cite{1,3, 12}, the impedance matrix must be recomputed for every successive frequency, limiting the applicability of the method to relatively small problems.

A first solution to improve the speed of the impedance matrix computation is to interpolate it w.r.t. frequency \cite{4}. However, the interpolation in the complex plane requires a 2D grid of sampling points in order to get accurate results. An alternative approach was proposed in \cite{5} which consists in reformulating the impedance matrix as a weighted sum of purely geometrical reaction integrals. These integrals can be computed once and for all. Then, the value of the impedance matrix for any frequency and material parameters can be obtained rapidly by summing these terms. The great advantage of this technique is that it relies on the Taylor expansion of a complex exponential. This Taylor expansion can be straightforwardly extended to the complex plane, providing accurate results using a 1D set of geometrical terms. In this paper, we use the latter technique to rapidly compute the impedance matrix for any complex frequency. Combined with an eigenmode solver, this greatly reduces the computation time required to determine the resonant frequencies of complex structures. 

The remainder of this paper is organized as follow. First, in Section 2, principle of resonant frequencies determination with the MoM are recalled. In Section 3, we remind the working principle of the FMIR-MoM and the way it can be used to rapidly compute the impedance matrix for several frequencies. In Section 4, we combine the rapid matrix computation of Section 3 with the method described in Section 2 to study the resonances of a plasmonic nanowire. It is shown that the time required to study the resonant frequencies of a relatively complex structure is greatly accelerated using the Geometrical MoM.

\section{Method of Moments formulation}
The Method of Moments (MoM) is based on Surface Integral Equations (SIE) \cite{2}. Using the equivalence principle, the fields scattered by a structure under a given illumination can be expressed using equivalent currents on the surface of the structure. Therefore, in order to find the total fields radiated or scattered by a structure, one only needs to determine these unknown equivalent currents. These currents can be found by enforcing the proper boundary conditions on the interfaces between different media. 

In a MoM scheme, the surface is first discretized using a set of vector basis functions $\mathbf{f}_{B,i}(\mathbf{r})$ to expand the unknown electric and magnetic currents $J_\text{eq}$ and $M_\text{eq}$ at any point $\mathbf{r}$ of the surface:
\begin{equation}
\mathbf{J}_\text{eq}(\mathbf{r}) \simeq \sum_i x^J_i \mathbf{f}_{B,i}(\mathbf{r})
\end{equation}
\begin{equation}
\mathbf{M}_\text{eq}(\mathbf{r}) \simeq \sum_i x^M_i \mathbf{f}_{B,i}(\mathbf{r})
\end{equation}
where $\mathbf{J}_\text{eq}(\mathbf{r})$ and $\mathbf{M}_\text{eq}(\mathbf{r})$ correspond to the unknown equivalent electric and magnetic currents in point $\mathbf{r}$ of the surface, respectively. Then, in order to find the unknown coefficients $x^J_i$ and $x^M_i$, the boundary conditions are imposed in the weak form using a chosen set of Testing Functions (TF) $\mathbf{f}_{T,i}$ resulting in a system of equations:
\begin{equation}
\label{eq:01}
\underline{\underline{Z}} \mathbf{x} = \mathbf{b}
\end{equation}
with $\underline{\underline{Z}}$ the impedance matrix of the structure, $\mathbf{x}$ a vector containing the unknown coefficients and $\mathbf{b}$ a vector accounting for the field incident on the structure.

Solving the system of equations of \eqref{eq:01} for a given excitation vector, one can find the equivalent currents on the surface of the structure and thus compute the scattered fields.

The Method of Moments can be used to compute the resonant frequencies of a structure \cite{1, 3, 12}. Indeed, mathematically, a resonant mode corresponds to a homogeneous solution to the Maxwell's equations, \textit{i.e.} a solution that can exist without an excitation. Looking at Equation \eqref{eq:01}, it means that a non-trivial solution $\mathbf{x}$ is possible even with a zero excitation ($\mathbf{b} = 0$). It is possible only if the rank of the impedance matrix is deficient. Therefore, determining the resonant frequencies of the structure amounts to determining the frequencies for which the rank of the impedance matrix is deficient. This deficiency can be estimated in several ways, either looking at the determinant of the matrix \cite{1} or any assimilated value \cite{3}, its smallest eigenvalue \cite{9} or its conditioning number \cite{10}.

The most straightforward way to determine these resonant frequencies is to compute one of the previous metrics of $\underline{\underline{Z}}$ for several complex frequencies to find valleys or peaks centred around the resonance \cite{10}. More evolved techniques have been proposed to limit the number of sampling points needed to find resonance, either by computing the determinant on a closed contour to determine the number of resonant frequencies enclosed \cite{3,9} or to converge iteratively to the closest one \cite{9}. However, for each new frequency, the impedance matrix needs to be recomputed, which is generally the bottleneck of such methods for structures of moderate size (few thousands of unknowns).

\section{Geometrical MoM}
Each entry $Z_{ij}$ of the impedance matrix corresponds to the interactions between BF $j$ and TF $i$. While the exact expression of the impedance matrix entry depends on the boundary conditions imposed, it physically corresponds to the fields induced on the TF by the currents on the BF. Thus, the current distribution on the BF must first be convolved with the Green's function and then integrated on the TF, leading to the evaluation of a 4D integral. If we note ${E}_{ij}^J$ and $H_{ij}^J$ the electric and magnetic fields generated on TF $i$ by the electric current on BF $j$, it reads \cite{2}
\begin{align}
\label{eq:02}
E_{ij}^J &= -\dfrac{j \eta}{4\pi k} \iint_{S'} \iint_S \Big(k^2 \mathbf{f}_{B,j}(\mathbf{r}) \cdot \mathbf{f}_{T,i}(\mathbf{r}') \\
&- \big(\nabla \cdot \mathbf{f}_{B,j}(\mathbf{r}) \big) \big(\nabla' \cdot \mathbf{f}_{T,i}(\mathbf{r}')\big) \Big) \dfrac{\exp(-jkR)}{R} dS'(\mathbf{r}') dS(\mathbf{r}),\nonumber
\end{align}
\begin{align}
\label{eq:03}
H_{ij}^J = -\dfrac{1}{4\pi} \dashiint_{S'} \iint_S& \nabla' \Bigg( \dfrac{\exp(-jkR)}{R}\Bigg) \\
& \times \mathbf{f}_{B,j}(\rpos) \cdot \mathbf{f}_{T,i}(\rposp) dS'(\rposp) dS(\rpos), \nonumber
\end{align}
with $\eta = \sqrt{\mu/\varepsilon}$ the impedance of the medium, $k = \sqrt{\omega^2 \epsilon \mu}$ the wavenumber, $\nabla$ and $\nabla'$ the derivative operator applied to the $\rpos$ and $\rposp$ coordinates, respectively, $S$ and $S'$ the surface of the BF and TF, respectively, $R = |\rposp -\rpos|$ the distance between the source point $\rpos$ and the image point $\rposp$ and $\varepsilon$ and $\mu$ the permittivity and permeability of the medium. The symbol $\dashiint$ denotes a Cauchy principal value integration (\textit{i.e.} an integration over the whole TF except for the point $\rposp = \rpos$). From \eqref{eq:02} and \eqref{eq:03}, the electric and magnetic fields  ${E}_{ij}^M$ and $H_{ij}^M$ generated on TF $i$ by magnetic current on BF $j$ can be computed straightforwardly considering that
\begin{equation}
{E}_{ij}^M = - H_{ij}^J,
\end{equation}
\begin{equation}
H_{ij}^M = E_{ij}^J/\eta^2.
\end{equation}

For each new frequency, the Green's function changes, such that the whole 4D integral must be evaluated again.

In the Geometrical MoM (GMoM), stemming from FMIR-MoM (Frequency and Material Independent Reaction) \cite{5}, the interaction between a pair of BF and TF is expressed as a weighted sum of purely geometrical 4D integrals. First, the geometrical terms are computed. Then, the impedance matrix can be obtained for any frequency and material parameters for practically no cost by simply summing up the geometrical terms.

To remove the frequency dependence of the integrand, the exponential term of \eqref{eq:02} and \eqref{eq:03} is expanded using its Taylor series w.r.t. the variable $R$ around a mean value $R_0$ corresponding to the mean distance between the BF and TF:
\begin{equation}
\label{eq:07}
\exp(-jkR) \simeq \exp(-jkR_0)\sum_{p=0}^N \dfrac{\big(-jk (R-R_0)\big)^p}{p!}.
\end{equation}
Using this expression and factoring the wavenumber out of the integral, we obtain:
\begin{equation}
E_{ij}^J \simeq -\dfrac{j \eta}{4\pi k} \exp(-j k R_0) \sum_{p=0}^N \dfrac{(-jk)^p}{p!} \big(A_p - k^2 B_p \big), 
\end{equation}
\begin{equation}
H_{ij}^J \simeq -\dfrac{\exp(-jkR_0)}{4\pi} \sum_{p=0}^N \dfrac{(-jk)^p}{p!} C_p,
\end{equation}
with $A_p$, $B_p$ and $C_p$ the purely geometrical terms:
\begin{align}
\label{eq:04}
A_p &= \iint_{S'} \iint_S \dfrac{(R-R_0)^p}{R} \\
&\hspace{2cm}\times \mathbf{f}_{B,j}(\rpos) \cdot \mathbf{f}_{T,i}(\rposp) dS'(\rposp) dS(\rpos) \nonumber
\\
\label{eq:05}
B_p &= \iint_{S'} \iint_S  \dfrac{(R-R_0)^p}{R} \big(\nabla \cdot \mathbf{f}_{B,j}(\rpos)\big) \\
&\hspace{2cm}\times \big(\nabla' \cdot \mathbf{f}_{T,i}(\rposp)\big) dS'(\rposp) dS(\rpos)\nonumber
\\
\label{eq:06}
C_p &= \dashiint_{S'} \iint_S \Bigg(\dfrac{(R-R_0)^p}{R^3} - p \dfrac{(R-R_0)^{p-1}}{R^2} \Bigg)\\ 
& \hspace{1.5cm}\times(\rposp-\rpos) \times \mathbf{f}_{B,j}(\rpos) \cdot \mathbf{f}_{T,i}(\rposp) dS'(\rposp) dS(\rpos) \nonumber
\end{align}
This method presents many advantages. First, the possibly singular integrals of \eqref{eq:04}, \eqref{eq:05} and \eqref{eq:06} admit a closed-form expression \cite{6}, providing a great precision at moderate computational cost. Second, it can be noticed that these integrands are purely real, lowering the memory and computational requirements by a factor of (at least) two. Third, since the expansion \eqref{eq:07} is valid in the complex plane, it can be used to evaluate the impedance matrix for complex frequencies. Fourth, no assumption is made on the value of $\varepsilon$ or $\mu$ so that this method accommodates easily for dispersive materials. Last, since the truncated sums of \eqref{eq:04}, \eqref{eq:05} and \eqref{eq:06} are corresponding to truncated Taylor expansions of the exponential of \eqref{eq:07} over distances corresponding to the electrical size of the BF and TF, few terms are required to reach a great accuracy. More information about the convergence of the method can be found in \cite{5}.

\section{Numerical example}
\begin{figure}
\center
\includegraphics[width = 8cm]{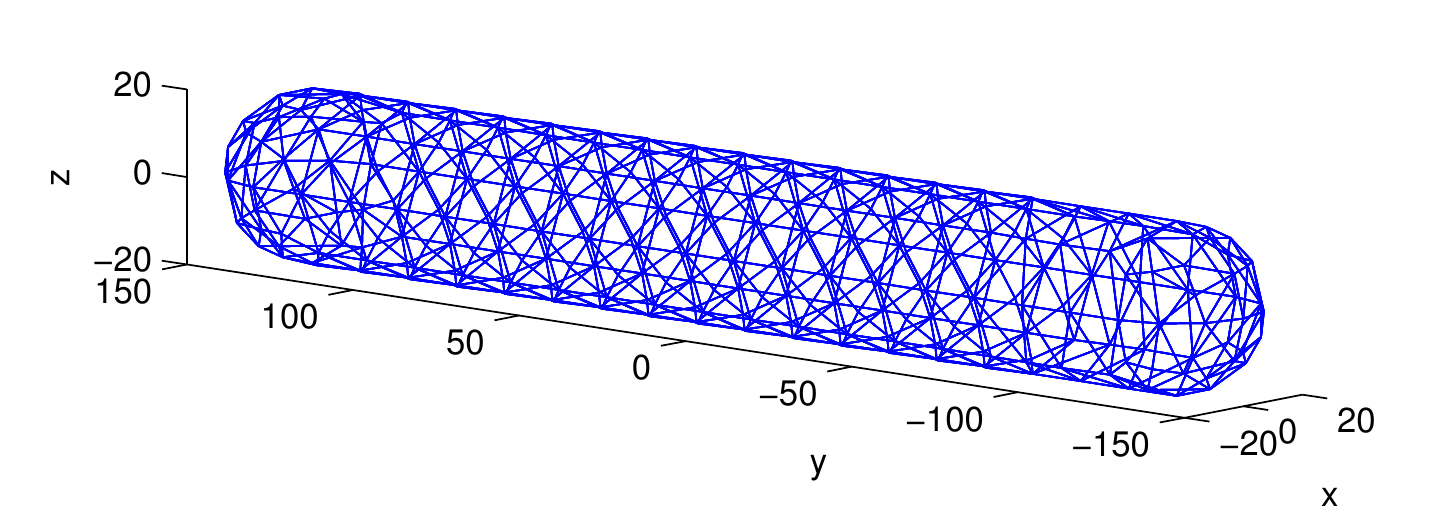}
\caption{Mesh used to discretize the surface of the wire.}
\label{fig:00}
\end{figure}

In order to validate our technique, we reproduced the results of Fig. 1 of \cite{1}. We simulated a 300 nm long free-standing gold nanowire with a diameter of 40 nm. The complex permittivity of gold was taken from \cite{8} considering only the real part of the frequency. Since we did not use any technique requiring the holomorphy of the impedance matrix with respect to the frequency \cite{3,9}, this simple extrapolation, which is not analytical, worked fine. The wire is discretized using 969 Rao-Wilton-Glisson (RWG) \cite{11} basis functions, leading to a total of 1938 unknowns. The structure can be seen in Fig. \ref{fig:00}.

The rank of the impedance matrix was evaluated for 100x100 frequencies whose real parts $f_r$ were regularly sampled between 200 and 500 THz and whose imaginary parts $f_i$ were regularly sampled between $2$ and $200$ THz. Since the amplitude of the blocks of the impedance matrix corresponding to the magnetic currents are known to be smaller than the blocks corresponding to electric currents by a factor $\eta$, we rescaled the impedance matrix accordingly by a factor corresponding to the mean impedance between air and gold. We did the same for the blocks related to electric and magnetic fields. In this way, the condition number of the matrix is improved and variations due to a resonance are more visible. Hereafter, we will refer to this matrix as the ``rescaled" impedance matrix.

\begin{figure}
\center
\includegraphics[width = 8cm]{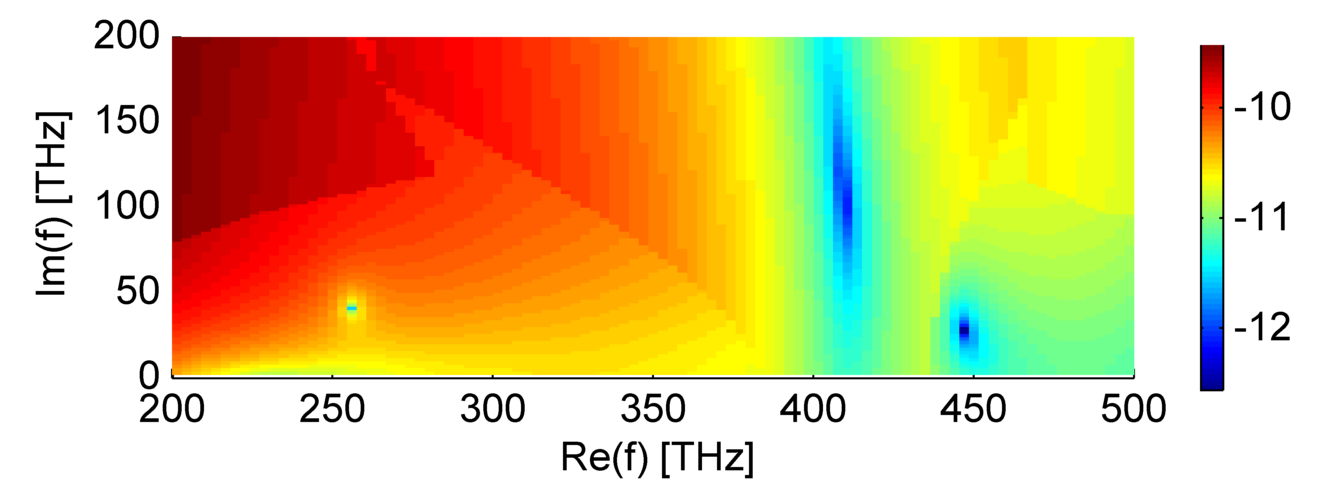}\\
(a) \\
\includegraphics[width = 8cm]{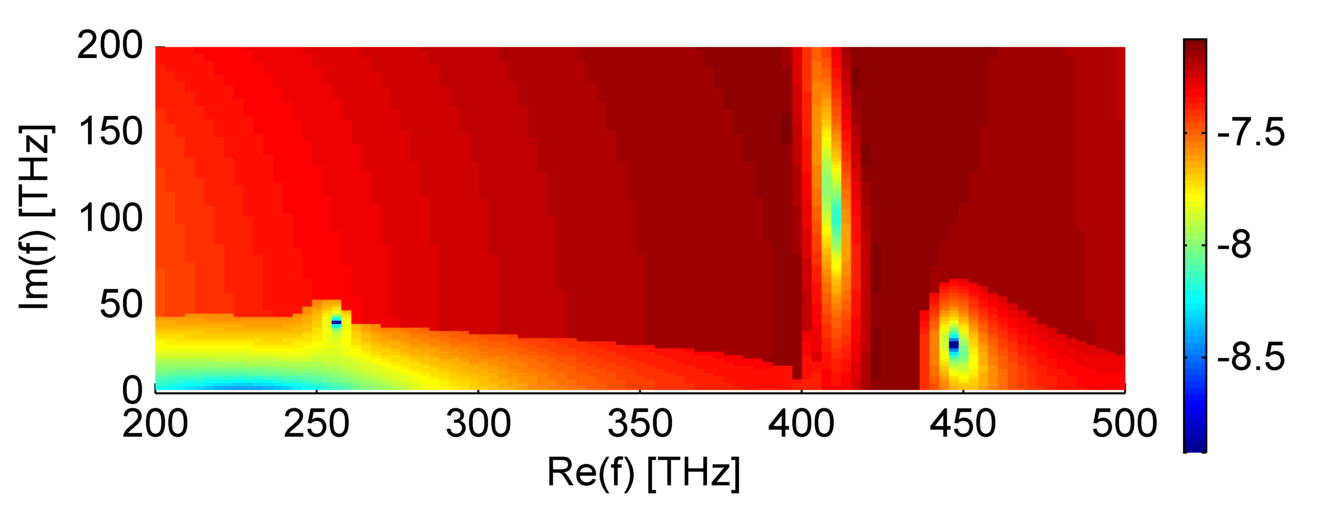}\\
(b)
\label{fig:01}
\caption{Map of the inverse condition number of (a) the impedance matrix and (b) the rescaled impedance matrix as a function of the complex frequency. The $\log_{10}$ of the value is displayed for better readability.}
\end{figure}

To get an indication on the rank of the impedance matrix, we used the {\tt rcond} function in Matlab language, which provides a fast estimate of the inverse of the condition number of the matrix. In that way, the closer the resonance, the smaller the result. We mapped the inverse condition number for both the original and the rescaled impedance matrix. The results obtained can be seen in Figure \ref{fig:01}. First, it can be noticed that in both cases, there are some discontinuities in the condition number map. We attribute these discontinuities to the {\tt rcond} algorithm of Matlab, which probably includes some branching statement whose output depends on the matrix under study. On both graphs, we can clearly see two very localized resonance around $256+40j$ and $445 + 26j$ THz. The first resonance matches the results of \cite{1} pretty well, while the second one seems to underestimate the imaginary part. It may be due to the difference in the way the permittivity of gold was extrapolated to the complex plane. We can also see a third and more diffuse resonance appearing around $410+102j$ THz that does not seem to be present in \cite{1} but could be hidden under the inset of the original figure.

Concerning the computation time, the simulation was performed on a laptop with an Intel i3-4000M processor. The computation of the geometrical terms took 450 seconds for 6 terms of the expansion. Then, for each frequency point, it took around half a second to compute the value of the impedance matrix and one second to evaluate its inverse condition number. 

It is important to note that, in this paper, we focused on the fast matrix computation for complex frequencies. However, the technique presented does not hinder the use of specialized iterative techniques such as those proposed in \cite{9}, which could dramatically reduce the number of complex-frequency samples needed.

\section{Conclusion}
In this paper, we proposed to combine the Geometric Method of Moments with an eigenmode computation algorithm to rapidly obtain the resonant frequencies of a structure. The Geometrical MoM is based on the precomputation of few purely geometrical reaction integrals which can be later combined to obtain the impedance matrix of the structure studied at any complex frequency for practically no cost. We validated the technique by reproducing available results in the literature. It is shown that, once the geometrical terms have been computed, the evaluation of the impedance matrix at any frequency is accelerated by several orders of magnitude.

\section*{Acknowledgment}
This project has received funding from the European Union Horizon 2020 research and innovation program under grant agreement No. 736937.

\end{document}